Experiments on the Photophoretic Motion of Chondrules and Dust Aggregates – Indications for the Transport of Matter in Protoplanetary Disks

Gerhard Wurm[1], Jens Teiser[1], Addi Bischoff[2], Henning Haack[3], Julia Roszjar[2]

[1]Faculty of Physics
University Duisburg-Essen
Lotharstr. 1
47057 Duisburg
Germany
e-mail: gerhard.wurm@uni-due.de
Tel.: +49 203 379 1641
Fax.: +49 203 379 1965

[2]Institut für Planetologie
Wilhelm-Klemm-Str. 10
48149 Münster
Germany

[3] Centre for Star and Planet formation
Natural History Museum of Denmark
University of Copenhagen
Øster Voldgade 5-7
DK-1350 Copenhagen K
e-mail: hh@snm.ku.dk
Tel.: +45 35 32 23 67
Denmark

Abstract

In a set of 16 drop tower experiments the motion of sub-mm to mm-sized particles under microgravity was observed. Illumination by a halogen lamp induced acceleration of the particles due to photophoresis. Photophoresis on dust-free chondrules, on chondrules, glass spheres and metal spheres covered with SiC dust and on pure SiC dust aggregates was studied. This is the first time that photophoretic motion of mm-sized particles has been studied experimentally. The absolute values for the photophoretic force are consistent with theoretical expectations for spherical particles. The strength of the photophoretic force varies for chondrules, dust covered particles and pure dust from low to strong, respectively. The measurements support the idea that photophoresis in the early Solar System can be efficient to transport solid particles outward.

Keywords: Disks; Experimental Techniques; Origin, Solar System; Meteorites; Solar Nebula

1. Introduction

It is well accepted that planetary systems form in protoplanetary disks. Protoplanetary disks consist of gas and solid particles surrounding a forming star. The interaction between solid particles and gas is important for the dynamics and short term evolution of the solid fraction within the disk as long as bodies are much smaller than 1 km, i.e. if they are not yet planetesimals (Weidenschilling, 1977). Lately, a growing number of young disks, often called transitional disks, have been observed which have almost dust free, optically thin inner regions but dense dust/gas disks further out (Sicilia-Aguilar et al., 2008; Calvet et al., 2002). At least some of these inner holes still contain significant amounts of gas (Najita et al., 2007). Especially the transitional disks emphasize that there are conditions in early protoplanetary disks where solids are present in a gaseous environment while they are illuminated by a directed light source.

It is well known that gas drag determines the relative motion between solids and gas. Relative velocities between dust particles are small, initially mm/s, and dust particles can collide gently, stick together and grow to larger aggregates (Blum and Wurm, 2008). Gas drag imposed on solids by viscous evolution of the disk has been proposed to lead to a net outward motion of solids under certain conditions (Takeuchi and Lin, 2002; Ciesla, 2007). In general radial transport is evident in the Solar System as high temperature minerals supposed to be formed close to the sun are found in comets supposed to be formed beyond 10 AU from the sun (Zolensky et al., 2006; Horner et al., 2007). Beyond the viscous transport, the most popular transport mechanisms which have been proposed to date are turbulence (Hogan and Cuzzi, 2001), or x-winds and stellar outflows (Skinner, 1990; Shu et al., 1996; Liffman and Brown, 1996). This is depicted in fig. 1. For dust size particles radiation pressure can also provide transport (Klacka and Kocifaj, 2001; Saija et al., 2003; Vinkovic, 2009). However, it has only recently been considered that the interaction between solids and gas due to the photophoretic effect can be equally important (Krauss and Wurm, 2005; Wurm and Krauss, 2006; Wurm and Haack, 2009).

If a temperature gradient exists across a solid particle thermal creep or interaction with individual gas molecules can lead to an efficient momentum transfer between particle and gas. If the temperature gradient is induced by light, the effect is called photophoresis (from the greek: transmission by light). Photophoresis is long known and has been described theoretically and experimentally by a number of authors (Fresnel, 1825; Crookes, 1874; Hidy and Brock, 1967; Tong, 1975; Pluchino, 1983; Rohatschek, 1995; Tehranian et al., 2001; Beresnev et al., 2003; Steinbach et al., 2004; Cheremisin et al., 2005; Wurm et al., 2008; Teiser et al., 2008; Jovanovic, 2009).

At low gas pressure, when solid particles can be considered to be in the free molecular dynamics regime, photophoresis might be understood as follows. Gas molecules continuously impinge the surface of a particle. A certain fraction – described by the thermal accommodation coefficient – sticks to the surface, before it is rejected again after a short time. At constant gas temperature the average momentum transferred to the particle by these gas molecules balances. However, the kinetic energy and momentum carried by the gas molecules upon rejection is determined by the local surface temperature of the particle. If two sides of a particle are at different temperatures then the gas molecules on the warmer side carry more momentum than the gas molecules rejected from the colder side. In total more momentum is transferred to the gas molecules leaving the warmer side. The particle has to balance this momentum transfer and is subject to a force directed from the the warm to the cold side – photophoresis.

In protoplanetary disks the conditions for photophoresis are given in different places and times and can transport particles efficiently. The transitional disks mentioned before are one obvious setting. Several other possibilities have been worked out in a number of recent publications by now (Krauss and Wurm, 2005; Wurm and Krauss, 2006; Mousis et al., 2007; Krauss et al., 2007; Herrmann and Krivov, 2007; Takeuchi and Krauss, 2008; Wurm and Haack, 2009). Possible ways of transport by photophoresis are indicated in fig. 1. From the inner disk, particles might be transported outward either in the midplane at later times when the total disk is transparent or over the surface at all times. Particles might be concentrated and "size" sorted in certain regions especially at the inner edge, in the asteroid belt or in the Kuiper belt. The latter is e.g. supported by models which consider the formation of the resonant population of objects in a gaseous disk (Griv and Jiang, 2009) or the formation of dust rings in a gaseous environment (Klahr and Lin, 2001).

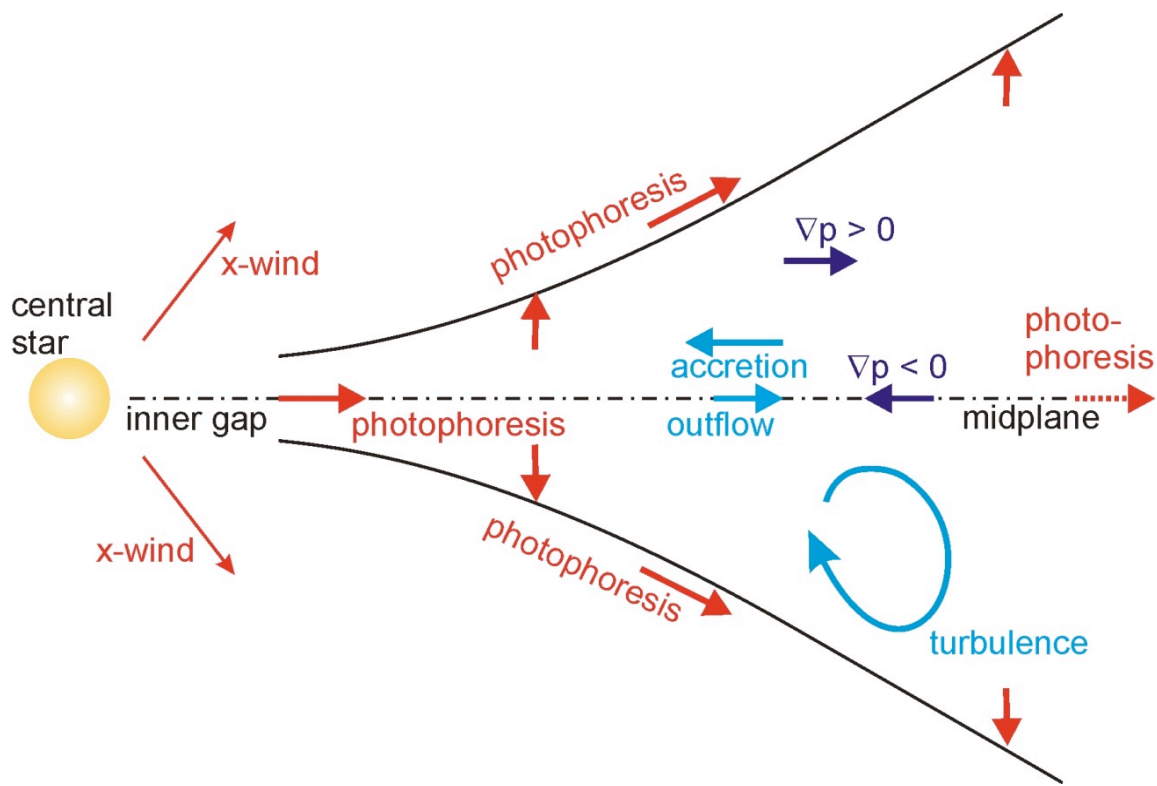

Fig. 1 Possible ways for radial transport of solids in protoplanetary disks. The sketch comprises different evolutionary stages of the disk. Vertical photophoresis is induced by thermal radiation from the disk. Photophoresis far out in the midplane is only possible at late stages when the disk is optically thin. For small dust grains radiation pressure acts in a similar way as photophoresis but is much weaker for large particles.

Photophoresis can transport dust aggregates to several tens of AU at least (Krauss and Wurm, 2005).

Photophoresis depends linearly on the radiative flux (see below). The photophoretic force therefore decreases with distance to the star. It might be worth considering that for full stellar radiation, the flux has the same distance dependence ($1/R^2$) as a star's gravitational attraction. However, in addition, due to the gas pressure dependence of photophoresis and extinction of starlight the force usually decreases faster than gravity with distance to a star. This allows the formation of equilibrium positions (Krauss and Wurm, 2005; Wurm and Krauss 2006).

A detailed treatment of the photophoretic force is given in section 2. In general it is worth noting that dust aggregates typically have much lower thermal conductivity than individual monolithic bodies of the same material due to their porous nature. Therefore, the temperature gradient established by illumination is stronger for an aggregate as for an individual particle, which results in a larger photophoretic force, respectively. Also the optical properties are important. The temperature gradient is depending on where radiation is absorbed within a particle. In extreme cases, radiation might be absorbed at the back of a particle. An artificial particle of that kind would be a glass particle painted black at the back side. In that case the temperature gradient is reversed compared to particles absorbing radiation at the front side and the particle moves toward the light source, which is called negative photophoresis. This is to emphasize that, while assumptions might be made for the photophoretic force on a given kind of particle, in detail the photophoretic strength strongly depends on the make-up of the particle.

The special protoplanetary disk assigned to the early days of our Solar System is often referred to as the Solar Nebula. Some more detailed information can be gained from the study of Solar System bodies. The most primitive meteorites (chondrites) contain up to mm-size particles called chondrules. Chondrules are the most abundant objects in many chondrite groups. In a general sense chondrules can be defined to be objects formed as isolated droplets of molten or partially molten material (Grossman et al., 1988). The presence of a glassy or cryptocrystalline mesostasis and an igneous texture are the strongest indications of a formation process by melting. The abundances and sizes of chondrules in different chondrite groups vary. Chondrules comprise a major fraction of the rocks in ordinary chondrites, but occupy smaller volumes in carbonaceous, enstatite (for detail see: e.g., Grossman et al., 1988) and Rumuruti chondrites (Bischoff, 2000). Evidence for size sorting has been discussed (Grossman et al., 1988 and references therein). Also, small mineralogical and chemical differences exist between chondrules in a chondrite group and between different chondrite groups. Most chondrules are dominated by olivine and low-Ca pyroxene and less abundant feldspathic material within the mesostasis. On average they are depleted in siderophile and chalcophile elements. Photophoretic transport might play a role in explaining part of the different properties of chondrites and chondrules (e.g. size sorting). Therefore, first measurement of the photophoretic strength acting on chondrules and related particles were carried out. The objective of this paper is to give first quantitative values for the photophoretic strength of mm-size particles. Section 2 presents the basic relations of photophoresis. Section 3 considers the experimental setup and particle samples used in the drop tower. Section 4 presents the data and data analysis. Section 5 remarks on the role of particle rotation. Section 6 summarizes the paper and puts the results in context with applications in protoplanetary disks and the Solar Nebula.

2. Photophoresis

The observation of photophoresis dates back at least to Fresnel (1825). Especially at high pressure photophoresis is not to be confused with a pressure rise due to heating on the warm particle side according to the ideal gas law ($p \sim T$). While the pressure would initially increase upon illumination and heating, any pressure difference is rapidly equilibrated in an open system with no walls. Photophoresis on the other side is a non-equilibrium phenomenon which in principle can be described by gas kinetics.

An explanation for photophoresis at low pressure has been given above. At high gas pressure, the gas is locally heated by the particle surface and gas is transported in a thin layer at the surface of the particle from the cold to the warm side (thermal creep) to which the particle reacts with a photophoretic force.

Rohatschek (1995) found a semi-analytical expression for the photophoretic force

$$F_{Ph} = \frac{2F_{max}}{\frac{p}{p_{max}} + \frac{p_{max}}{p}} \qquad \text{Eq. 1}$$

$$p_{max} = \frac{\eta}{a}\sqrt{12\frac{RT}{\alpha\mu}}$$

$$F_{max} = \left(\pi\sqrt{\frac{R}{3\mu T}}\eta a^2 I\right)\left(\sqrt{\alpha}\frac{J}{k}\right)$$

Within this equation parameters which are known or have been measured are the gas constant, $R$, the particle radius, $a$, the radiative flux $I$, the gas viscosity, $\eta$, the gas temperature, $T$, the gas pressure, $p$, and the molar mass of the gas, $\mu$. Rohatschek (1995) also had a thermal creep parameter which is close to 1 and which is assumed to be one in eq. 1, to be consistent with the approximation for low pressures given below in eq. 2. The pressure dependence is well visualized in eq. 1. There is a linear increase in the photophoretic strength with pressure at low pressure and a 1/p decrease at large pressure. The maximum photophoretic force is given for the intermediate pressure range or Knudsen numbers around Kn ~ 1. The Knudsen number is defined as Kn=λ/a, where λ is the mean free path of the gas molecules.

Particle parameters, which enter in eq. 1, and are not easily determined for small irregular or non-homogeneous particles are $k, J, \alpha$. The thermal conductivity, $k$, in the underlying model of eq. 1 is assumed to be a scalar value. In reality a complex particle will not be homogeneous with respect to $k$, nor will it necessarily be isotropic, i.e. heat conduction might depend on the direction. The thermal accommodation coefficient α refers to interaction of gas molecules with a particle's surface. In a simple picture of gas in the free molecular regime molecules are either reflected specular on a surface or are adsorbed on the surface for a certain time and re-ejected diffusely afterwards. If the fraction of specular reflected molecules is the same all over the surface of a particle this fraction does not contribute to any net momentum transfer between particle and gas. However, the adsorbed molecules are ejected with a velocity determined by the local surface temperature which leads to a net momentum transfer for a non-uniformly heated particle. This fraction is referred to as thermal accommodation coefficient α. It is often close to 1 (never larger) but in detail depends on the gas and the surface properties of the particle. The asymmetry parameter, $J$, summarizes the interaction of the particle with the radiation. The determination of $J$, in general, is a highly complex problem for itself. A particle which perfectly absorbs at the surface of the front side has $J$=0.5. If the pressure dependence (eq. 1) is measured accurately, α can be measured by determining the position of $p_{max}$. However, for large Knudsen numbers the photophoretic force is approximately (Beresnev et al., 1993)

$$F_{Ph} = \frac{\pi a^3 I p}{3T} b$$  Eq. 2

$$b = \frac{J\alpha}{k}$$

All unknown parameters then only occur within a single factor $J\alpha/k$, which is defined as $b$ here. The measurement of the photophoretic force at large Kn does not allow determining all three parameters individually. Eq. 2 and eq. 1 do not consider cooling by thermal radiation to be important for the temperature gradient. Also heat transfer by interaction with the surrounding gas is neglected. Both effects are included in a more rigorous treatment by Beresnev et al. (1993). However, these additional effects are negligible compared to heat conduction through the particle under the conditions relevant here.

3. Drop tower experiments

Using typical or accessible values in eq. 2 ($I = 10$ kW/m$^2$, $p = 0.1$ mbar, $k = 1$ W/mK, $J = 0.5$, $\alpha = 1$, $T = 300$ K) the ratio between photophoretic force and Earth gravity ($F_g=mg$) is on the order of $10^{-3}$, assuming a particle density of $\rho = 3000$ kg/m$^3$. This shows that gravity is dominating in Earth bound experiments. A set of 16 experiments at the drop tower Bremen was carried out, where microgravity conditions with a residual acceleration better than $10^{-5}g$ are given for 4.7s of free fall time. An acceleration of $10^{-2}$ m/s$^2$ is easily measurable then.
    Fig. 2 shows a sketch of the setup.

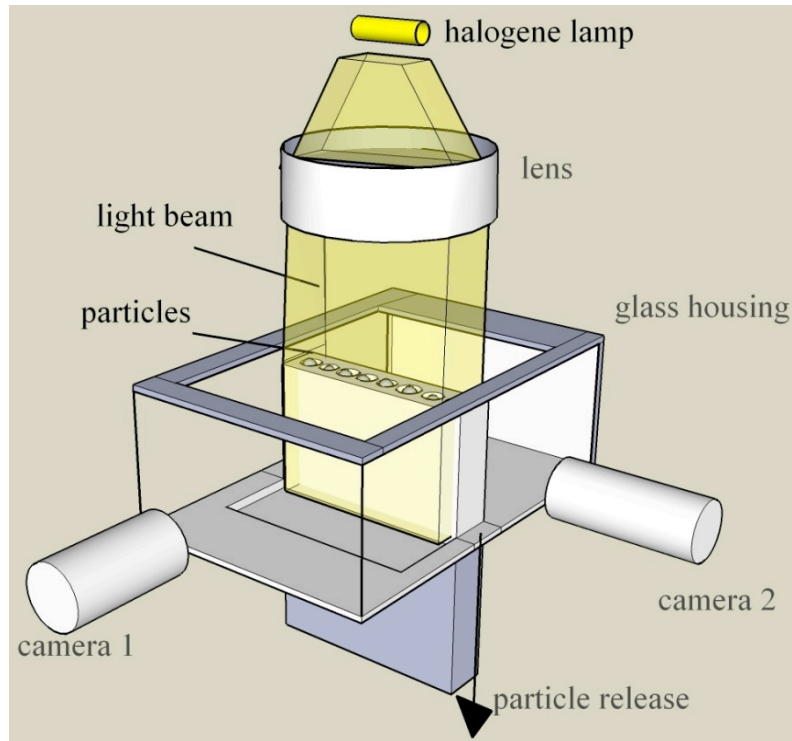

Fig. 2 Sketch of the drop tower experiment. A particle sample is released under microgravity in a low pressure gaseous environment. The motion induced by illumination with visible light is observed by two cameras.

The setup has three principle components, (a) the experiment chamber with the particle release mechanism and confinement area, (b) the illumination, and (c) the video observation. The latter is consisting of two cameras which observe the particle motion in two orthogonal directions. Camera 1, which is placed perpendicular to the particle array, has a clear view to all of the particles during the experiment. Examples of images from this camera are shown in fig. 3.

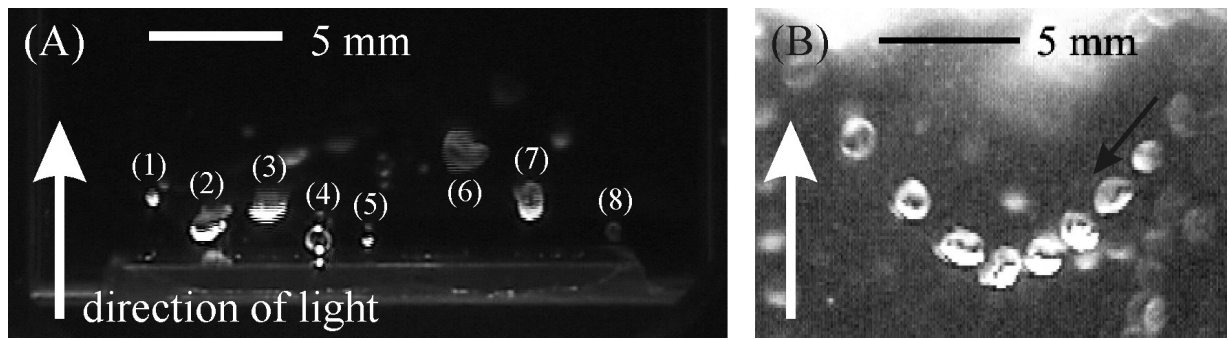

Fig 3 (a) Lateral view of particles from camera 1. Illumination is from bottom to top and marked by arrows. Illuminated chondrules (1 and 5), dust mantled glass (3 and 6) and steel (2) spheres, an illuminated glass sphere (4) and a darker (irregular) chondrule (8) not within the light sheath are visible. The fainter particle images are reflections. (b) A superposition of 10 images visualizing the acceleration of a dust mantled glass sphere within the light beam. The white arrow marks the direction of light. Due to bouncing at the glass wall the initial movement of the particle is directed along the black arrow.

The motion determined along the direction of light from camera 1 is used for the data analysis. Camera 2 is used for control purpose, e.g. to decide if a particle is illuminated or within the dark region or if a sudden change in particle motion is due to particle-wall collision.
    As illumination a halogen lamp was used, which was placed outside of the vacuum chamber. The radiation was focused by a condenser, also placed outside of the chamber, to a rectangular shaped cross section within the experiment chamber. The beam cross section at the initial position of the particles is visualized in fig 4 and is about 20 x 5 mm$^2$.

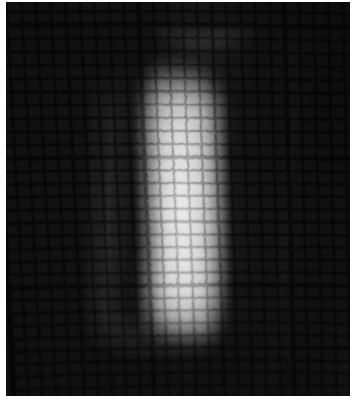

Fig. 4. Cross section of the the light beam on a sheet of mm-squares.

As the cross section is the result of a slightly defocused and tightly screwed hot wire, the beam profile is of relatively uniform rectangular shape. About 2 mm from the center line the intensity drops by about 40% before it reaches a steep cutoff.

As the beam is divergent, the cross section changes along the direction of light with distance. The particle is free to move about 25 mm along this direction. Therefore, the maximum light flux as a function of distance from the initial position of the particles was measured. As the particle, in principle, can move these 25 mm within the light beam some variation in the strength of the photophoretic force is induced. However, the variations are only about 10% in light intensity which is smaller than the uncertainty induced already by the variation over the beam cross section. We will use the typical (maximum) flux measured at the center of the spot at the initial particle position for the calculations below. This means that the photophoretic forces which are calculated are systematically underestimated and might be a factor of 2 higher.

A value of 25.5kW/m$^2$ was measured as light flux at 28V power supply. However, the power supply of the experiment provided only 25.3V for the halogen lamp during free fall. The total measured optical output integrating over the whole cross section is 1.69W at 25.3V and 1.89W at 28V which directly enters as 0.89 times the measured light flux available. Another reduction in available light flux is induced by the windows. The experiment chamber has one window sealing the vacuum chamber from the environment and one glass plate as part of the confining particle cage within the chamber. The difference between light flux with and without windows at different power supplies (halogen lamp temperatures, light flux) was measured and a linear dependence was found, i.e. the extinction by windows is not depending on the temperature of the lamp. In total the extinction by the windows leads to a reduction in the available light flux with respect to the measured light flux given above by a factor 0.83. As the light passes the condenser before, this also indicates that the fraction of thermal radiation (beyond 2μm) incident on the vacuum chamber and particles is not significant.

All together this leads to a maximum initial light flux of 18.8 kW/m$^2$ but with possible variations at later times down to 9.4 kW/m$^2$. The light source is considered as being restricted to the visible part of the spectrum not further specified and the beam is assumed to be unidirectional to first order, noting that with a condenser of 40 mm aperture and a working distance of ~150 mm angular variations up to ~8° are present.

Particles have to be confined within the beam cross section of 5 mm width for a significant time to be subject to photophoresis. Particles therefore initially have to be slower than a few mm/s. The initial movement of particles lifting off a surface at the onset of microgravity is usually larger as the tension release of the setup at the transition from gravity to microgravity accelerates the particles. The particles were therefore kept in place in the center of the light beam by being enclosed in small cavities which are pressed against the top glass plate of a glass cage within the vacuum chamber. The particles are released to be free moving ~0.1 s after the onset of microgravity. Collisional damping within the cavity has slowed down any initial movement then to a sufficiently small degree to keep a number of particles within the light beam over the course of the microgravity time.

Ten particles are observed in each experiment clearly separated and individually recognizable. The glass cage within the vacuum chamber has been used for different reasons. First, the top plate is used as counterpart to seal the cavities which enclose the particles until release. The particles are released by pulling the plunger with the cavities out of the glass cage. A glass bottom part then slides in position to close the cage. The glass cage then confines the motion of the particles to the observable volume which allows tracing trajectories for non-illuminated particles as well. The small volume compared to the larger vacuum chamber also eases the process of retrieving the particles after an experiment.

The main objective of the experiment was to quantify photophoresis on chondrules for the first time. Therefore, a number of chondrules were separated from the meteorite Bjurböle, which has been classified as L/LL4 chondrite. The chondrules were then treated by a grinding solution in a rotating drum until any residual matrix material was removed. Each chondrule was then photographed and the mass was determined. Images of some chondrules used are shown in fig. 5.

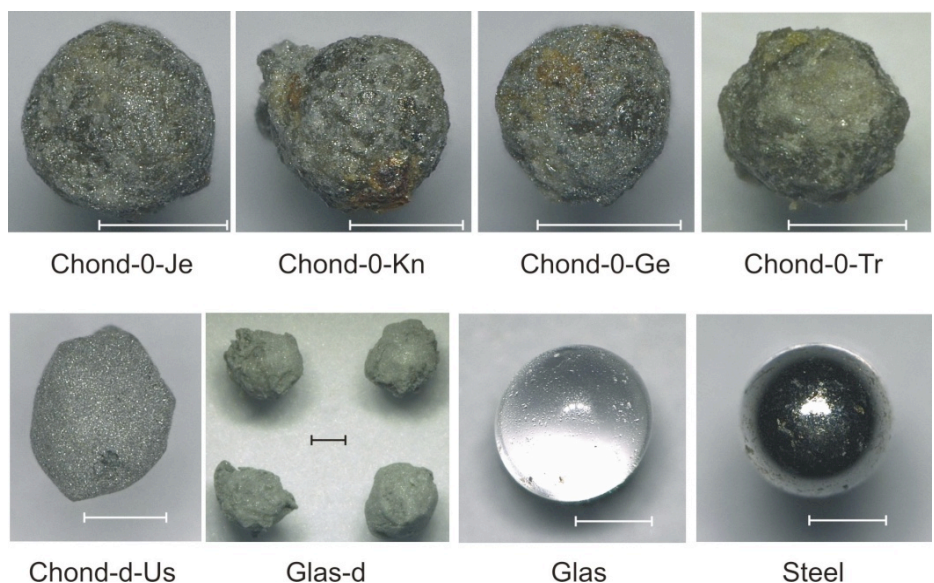

Fig. 5 Selection of used particles: chondrules, dust covered chondrule, dust covered glass spheres, glass sphere, and metal sphere. The scale bars are 500 μm each.

From the images, radii of the chondrules were determined. These are typically accurate to about 10% due to non-sphericity. Table 1 summarizes the values for the chondrules and other particles used within the experiments. All chondrules have individual features which allow their recognition and retrieval after an experiment.

Table 1: Properties and measured acceleration for the particles used. Negative accelerations are attractions by the window for particles in close proximity to the window. They are not considered in the further data analysis (see text). Particles immediately leaving the light sheath or sticking to the window are not listed. Dust aggregate masses are derived from the measured size.

| Particle name 0:bare, d: dust | Mass (mg) ± 1% | Diameter (mm) ±10% | Pressure (Pa) | Acceleration (mm/s²) |
|---|---|---|---|---|
| Chond-0-Je | 0.606 | 0.71 | 9 ± 1 | 1.242 ± 0.026 |
| | | | 12 ± 2 | 0.978 ± 0.016 |
| | | | 15 ± 1 | 1.020 ± 0.008 |
| | | | 0.41 ± 0.01 | 0.332 ± 0.008 |
| | | | 7 ± 1 | 2.078 ± 0.020 |
| | | | 4.7 ± 0.2 | 0.682 ± 0.010 |
| Chond-0-Ge | 0.309 | 0.59 | 7 ± 1 | 3.530 ± 0.130 |
| | | | 0.40 ± 0.01 | -2.752 ± 0.052 |
| | | | 12 ± 1 | -0.370 ± 0.082 |
| | | | 10 ± 1 | 1.246 ± 0.082 |
| Chond-0-Kn | 0.736 | 0.83 | 10 ± 1 | 1.280 ± 0.028 |
| | | | 0.41 ± 0.01 | 0.230 ± 0.020 |
| | | | 9 ± 1 | 1.686 ± 0.012 |
| Chond-0-Pa | 0.810 | 1.00 | 9 ± 1 | 1.750 ± 0.014 |
| | | | 15 ± 1 | 0.452 ± 0.010 |
| | | | 4.7 ± 0.2 | 9.260 ± 0.074 |
| Chond-0-Tr | 0.597 | 0.76 | 9 ± 1 | 3.020 ± 0.052 |
| | | | 12 ± 2 | 4.896 ± 0.068 |
| Chond-0-Ka | 0.608 | 0.77 | 0.41 ± 0.01 | 0.174 ± 0.008 |
| | | | 12 ± 2 | -2.318 ± 0.040 |
| Chond-0-Ch | 16.76 | 2.20 | 12 ± 1 | 0.710 ± 0.010 |
| | | | 7 ± 1 | 2.526 ± 0.030 |
| Chond-0-Frau | 0.101 | 0.43 | 4.7 ± 0.2 | 4.730 ± 0.024 |
| | | | 10 ± 1 | 1.760 ± 0.012 |
| Chond-0-Ju | 0.378 | 0.68 | 4.7 ± 0.2 | 1.306 ± 0.032 |
| Chond-0-He | 1.301 | 1.00 | 11 ± 1 | 2.278 ± 0.142 |
| Chond-0-Ja | 0.088 | 0.42 | 0.41 ± 0.01 | 0.436 ± 0.008 |
| Chond-0-Fro | 0.620 | 0.88 | 12 ± 2 | -1.606 ± 0.070 |
| Chond-0-Us | 0.735 | 0.81 | 12 ± 2 | -0.658 ± 0.068 |
| Chond-0-Vl | 0.078 | 0.15 | 0.41 ± 0.01 | 0.800 ± 0.010 |
| Chond-0-Vi | 0.508 | 0.68 | 0.41 ± 0.01 | 0.260 ± 0.006 |
| Chond-0-Li | 0.605 | 0.73 | 0.41 ± 0.01 | 0.288 ± 0.008 |
| Chond-0-Si | 0.116 | 0.43 | 7 ± 1 | 0.872 ± 0.076 |
| Chond-d-Us | 0.959 | 1.06 | 11 ± 1 | 46.794 ± 0.310 |
| | | | 12 ± 1 | 87.682 ± 1.782 |
| Chond-d-Pa | 1.272 | 1.22 | 12 ± 1 | 28.582 ± 0.336 |
| | | | 11 ± 1 | 23.684 ± 0.306 |
| | | | 0.40 ± 0.01 | -1.364 ± 0.056 |
| Metal-d-3 | 4.924 | 1.40 | 10 ± 1 | 2.916 ± 0.188 |
| | | | 11 ± 1 | 7.864 ± 0.080 |
| Metal-d-7 | 4.800 | 1.29 | 10 ± 1 | 7.142 ± 0.116 |
| | | | 11 ± 1 | 4.292 ± 0.286 |
| Glass-d-F | 1.737 | 1.42 | 10 ± 1 | 21.598 ± 0.490 |
| | | | 11 ± 1 | 36.778 ± 0.768 |

| Glass-d-D | 3.062 | 1.90 | 11 ± 1 | 55.720 ± 0.364 |
| | | | 0.40 ± 0.01 | 1.568 ± 0.106 |
| | | | 12 ± 1 | 61.700 ± 2.132 |
| Dust Aggregate 1 | 2.68 | 1.47 | 0.40 ± 0.01 | 1.490 ± 0.198 |
| Dust Aggregate 2 | 2.57 | 1.45 | 0.40 ± 0.01 | 21.916 ± 2.192 |
| Dust Aggregate 3 | 2.17 | 1.37 | 0.40 ± 0.01 | 21.916 ± 0.316 |
| Dust Aggregate 4 | 6.06 | 1.93 | 12 ± 1 | 176.232 ± 2.062 |
| Dust Aggregate 5 | 1.94 | 1.32 | 12 ± 1 | 132.316 ± 7.388 |
| Dust Aggregate 6 | 2.21 | 1.38 | 12 ± 1 | 200.780 ± 4.774 |

While keeping in mind that the chondrules might have been subject to some modification since their formation 4.5 billion years ago, these measurements give first values for the photophoretic strength on bare chondrules. However, chondrules might have been covered by dust in the solar nebula (Metzler et al., 1991, 1992; Metzler and Bischoff, 1996; Ormel et al., 2008). To simulate the strength of photophoresis on dust covered chondrules, a layer of SiC dust was glued to some chondrules, glass and metal spheres. Particles were covered by a thin layer of glue and were moved through the dust. The typical dust grain size of the dust is 10 μm. The thickness of the dust/glue layer varied between 20% and 40% of the coated particle size (see table 1). We are aware that this is by way not a representative dust cover for particles in the solar nebula. The glue likely increases the thermal conductivity as it filled pore space which would otherwise contribute to low thermal conductivity. However, covering particles with dust without glue did not prove stable enough to be used in the experiments. Nevertheless, this simulates a lower thermal conductivity of a dust rim and a change in the optical properties, e.g. absorption in a thin surface layer at the illuminated side of the particle. The mass of the dust covered particles was determined after the experiment campaign at the drop tower.

To follow along this line also pure SiC dust aggregates were used. Dust aggregates were not glued together but were stable enough to be used without further processing. However, they did not survive deceleration and masses could not be determined but were only estimated by means of the size, bulk density ($3217 kg/m^3$), and a filling factor assumed to be 0.5. We note that the filling factor is only an estimate and likely might vary. The highest filling factor possible is 0.64 for random close packing. The lowest value for which the particle is likely stable enough, not to be compacted further by the handling might be estimated to 0.33 (Blum and Wurm, 2008). Therefore, the individual values for the aggregate mass determined and therefore the photophoretic force uncertainties imposed are within a factor of 2. However, this does not influence the principle result that dust aggregates experience the strongest photophoretic force as seen below.

Furthermore, 1 mm glass spheres and steel spheres were used in the experiment. Both should show insignificant photophoretic acceleration compared to the other particles. Glass spheres do not absorb visible radiation and steel has a high thermal conductivity and albedo. These particles were used as test particles to show any other possible effects like thermophoresis, gas drag by residual gas motion or electrostatic forces which might be present.

## 4. Data analysis

From the video images particle tracks along the direction of light were determined. From the particle tracks the particle accelerations could be deduced. As seen in fig. 6, a chondrule is accelerated as long as it is illuminated. The track shows a linear motion if the particle moves in non-illuminated parts of the cage. A fitted parabola and linear motion are overplotted in fig. 6.

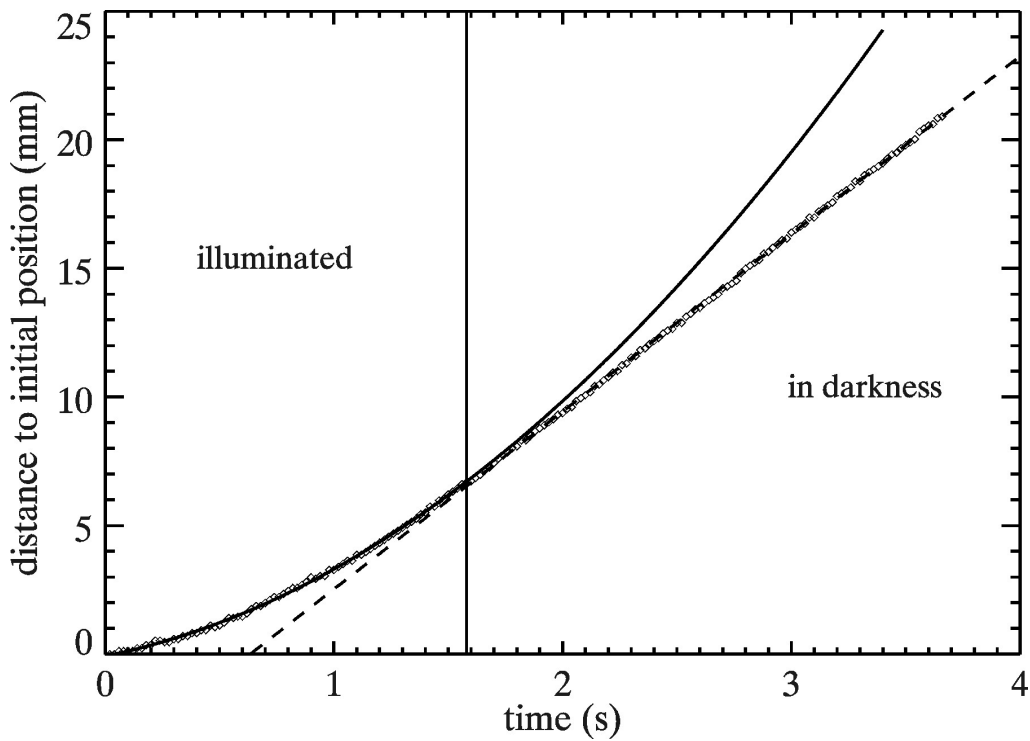

Fig. 6 Track of a chondrule (chond-0-Tr, 3.0 mm/s$^2$, table 1) crossing from the illuminated to the dark region of the experiment chamber. The symbols mark the position of the chondrule in each frame. A parabola is fitted to the illuminated part of the track and a straight line to the dark part.

Exceptions of linear motion and photophoretic positive acceleration are observed close to the glass walls of the confining cage, where particles can be attracted to the wall, most likely due to thermophoretic effects but we do not follow this aspect here. To avoid influence of such effects we did not analyze data of particles close to the walls.

In general all particles which are not within the illuminated area and which are in the center parts (not within 2 mm of the walls) of the chamber move on straight lines. The glass and metal spheres are not influenced by photophoresis and also move on straight lines with constant velocity within the illuminated area as expected. The fact that all particles move with constant velocity in the non-illuminated analyzed region confirms that only light induced effects and no gas drag, electrical charges or thermophoretic forces were measured by means of determining the acceleration. This is consistent with rough estimates of these forces.

At 10 Pa the gas grain friction time τ of a 1 mm particle is larger than 10 s (Blum et al. 1996). The acceleration due to a gas flow at constant velocity v would be a = v / τ. For a gas flow of 1 cm/s the acceleration gets comparable to the measured values. Blum et al. (1996) showed that at such low pressures any initial gas motion is damped within ~0.1 s in a much larger volume chamber and that residual velocities are two orders of magnitude smaller (~100 µm/s). Therefore, gas drag due to initial gas motion should not be observed. Electrostatic effects are not as easily estimated. In general, if charging of particles, e.g. due to triboelectric mechanisms occurs, the accelerations can be as large as the measured values. However, the charge distribution on the isolating glass cage would likely be inhomogeneous and the accelerations should strongly depend on the distance to specific charged spots. Also accelerations should not depend on particles being illuminated or not being illuminated. The latter is not what we see. We observe some acceleration close to the walls, which is not due to illumination. We cannot rule out that this is due to electrostatic effects but we only consider particle motion in regions of the chamber where no deviation from straight trajectories of particles is observed if particles are not illuminated. Therefore, our data are not biased by electrostatic effects. Thermophoretic effects might also play a role. If a temperature gradient within the chamber already exists, particles would experience a thermophoretic force. At low pressure (large Knudsen numbers) the thermophoretic acceleration is given as (Zheng 2002)

$$a_{th} = \frac{4\sqrt{\pi}\kappa_g}{5\pi\rho a\sqrt{2k_B T/m_g}} \frac{dT}{dx}$$   Eq. 3

With $\kappa_g$ = 0.01 W/Km being the thermal conductivity of the gas this is $a_{th}$= 9 $\Delta T$ mm/(s$^2$K) for a density of 3000 kg/m$^3$ and a particle radius of 0.5 mm at 300 K gas temperature. The temperature difference is assumed to be given over the extend of the glass housing of 25 mm. A temperature difference of 1 K over the length of the glass housing would lead to accelerations comparable or even larger to the measured accelerations. This is within a range that can occur in general so we took precaution to avoid thermal gradients. The experiment was set up to minimize temperature gradients. The glass cage was placed within a large vacuum chamber and no specific heat sources were present at any side except the light source itself. The light source was only switched on during microgravity to prevent a heating of the experiment setup. As thermal radiation was absorbed by the optical parts (condenser) outside of the glass cage only the visible light and only within the illuminated area was fed to the particles. Via thermal coupling to the particles and the particle mount the illuminated strip of the glass housing might have heated up during a fraction of a second but this would be a local heat source on the surface. The particle mount itself was removed from the glass housing which was then sealed by a glass plate which slides into place and prevents thermal radiation from the particle mount to influence the measurements. Therefore, in total thermal gradients along the experiment volume were kept to a minimum. As no temperature sensors were installed, we cannot quantify thermophoretic forces further. However, again any thermophoretic force would occur independently of particles being illuminated or not while free floating. We did not observe this. Ultimately, from this measured linear, constant motion of the glass and metal spheres as well as other particles moving in different non-illuminated regions we rule out thermophoresis as source for the measured accelerations.

Radiation pressure, which has similar effects as photophoresis, is not important. As detailed in Krauss and Wurm (2005) radiation pressure forces in the given parameter space are usually orders of magnitudes smaller than photophoretic forces. To quantify this here we consider the acceleration by radiation pressure as

$$a_{rad} = \frac{I}{c} \frac{3}{4\rho a}$$　　　　Eq. 4

This assumes perfect absorption over the cross section of the particle. The speed of light is *c*. For *a* = 0.5 mm, ρ = 3000 kg/m$^3$ and *I* = 18.8 kW/m$^2$ this is $a_{rad}$ = 0.008 mm/s$^2$. For most cases this is indeed two or three orders of magnitude less than the acceleration measured. Only in a few cases it is up to a few % of the measured acclerations (see table 1). In total, we therefore conclude that we measured photophoresis and no other forces.

A last thing to consider is a variation of the photophoretic strength with time if the temperature gradient along the particle is not yet in equilibrium. As it needs some time to get established, the photophoretic force might increase with time. A worst case estimate can be based on the typical timescales for conductive heat transfer through the particle

$$\tau_{heat} = \frac{\rho_d c_d a}{k}$$　　　　Eq. 5

where $\rho_d$ is the particle density, and $c_d$ is the heat capacity. This assumes that the surface temperatures are strongly related to the heat conduction through the particle. For a 1mm particle with *k*=1W/mK and assuming $\rho_d$=3000 kg/m$^3$ and $c_d$=1000 J/kg K we get τ=3s. This is comparable to the free fall time or experiment duration. It therefore has to be considered that chondrules are not fully equilibrated yet with respect to their temperature while being accelerated. In general the photophoretic strength should increase over time as we only turn on the light at the onset of microgravity. However, we did not see any change in the acceleration with time if we divide a trajectory in several parts. Within the uncertainties of the trajectory determination, particle tracks are consistent with a single acceleration only. Obviously the temperature gradient across the surface of the particles studied is fast enough to get close to equilibrium in a fraction of a second.

In one case a particle rotated with more than 6 Hz and a weaker photophoretic force was measured compared to similar particles (see below). In that case the temperature gradient might not be established fast enough. Sideward motions are also observed then, i.e. a particle is not heated homogeneously and temperature gradients are not necessarily directed away from the light (see below). We argue that the heat conduction time overestimates the time needed to establish a surface temperature close to the equilibrium case. If we e.g. consider a particle which would have a negligible small thermal conductivity, then the heat conduction time would be indefinitely large but the illuminated surface would – on the other side – instantly reach a temperature where thermal emission balances the absorption. Detailed heat transfer calculations are needed for more quantitative statements in the future.

In the following we only consider the linear motion away from the light source in view of the simple model given in eq. 1 and eq. 2. We assume that the measured acceleration is representative of an equilibrium condition, keeping in mind that a decrease of the photophoretic force due to rotation of the particle or reduced average light flux might be superimposed on this and that the exact value, depending on rotation and orientation, can differ for each experiment.

Fig. 7 shows the absolute photophoretic forces measured. As expected, due to the lower thermal conductivity compared to chondrules, photophoresis is stronger for the dust covered particles and strongest for dust aggregates.

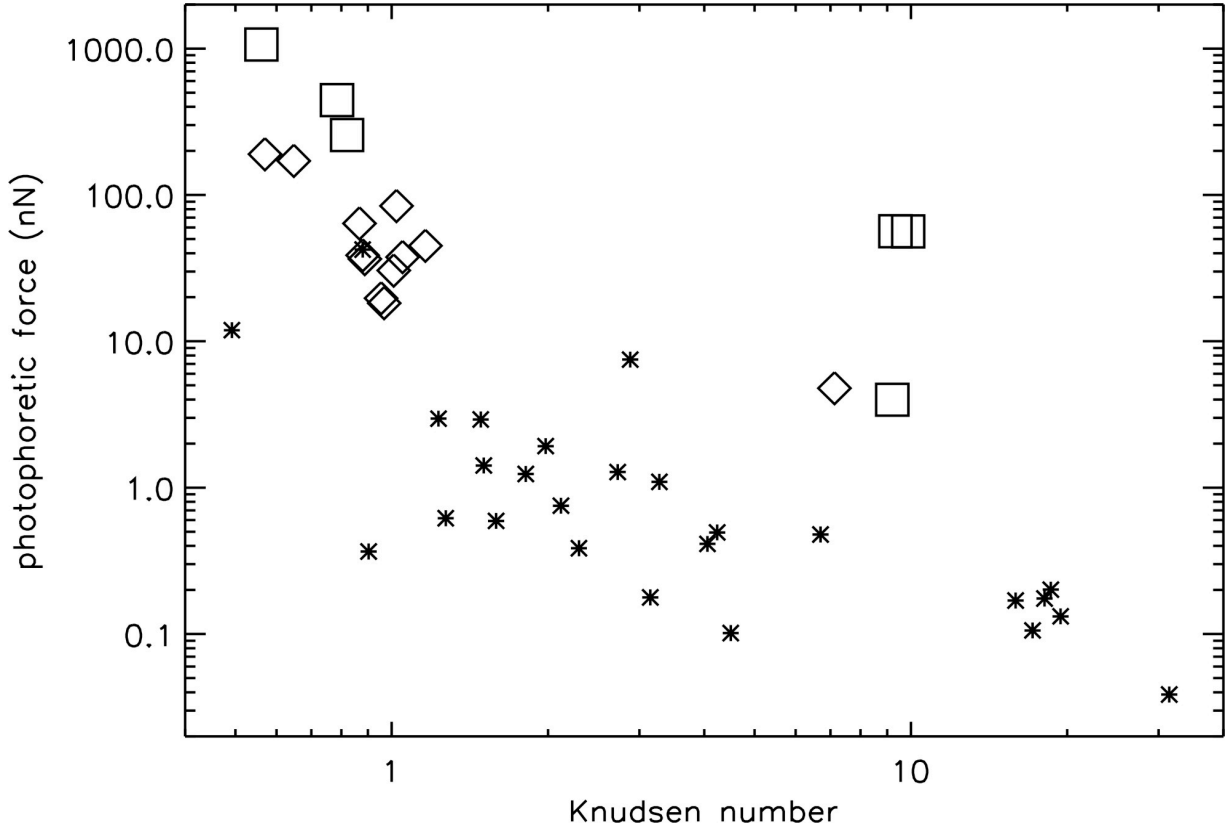

Fig. 7 Absolute photophoretic forces measured for chondrules (stars), dust mantled particles (diamonds) and dust aggregates (squares).

Fig. 8 has a similar structure as fig. 7 but shows the ratios between measured values and photophoretic forces $X=F_{meas} / F_{calc}$ for the different particle groups calculated by using eq. 1. As temperature we used 300K. For the calculations we assumed $k$=1W/mK, $J$=0.5, and $\alpha$=1 or $b_{calc}=kJ\alpha$=0.5W/mK. These values are standard assumptions, i.e. assuming perfect absorption at the front side, and perfect accommodation ($\alpha$=1). All 3 unknown individual parameters occur as product $b$ in eq. 2 and cannot be quantified individually within the accuracy of the given experiment. While we calculate $X$ using eq. 1 which is valid for all Knudsen numbers $X$ is also approximately giving $b_{meas} / b_{calc}$. Average values for the different particle groups are $X_c$=0.10 for chondrules, $X_{dp}$=0.57 for dust mantled particles (not specifying the core particle), and $X_d$=3.47 for dust aggregates.

(Fig. 8)

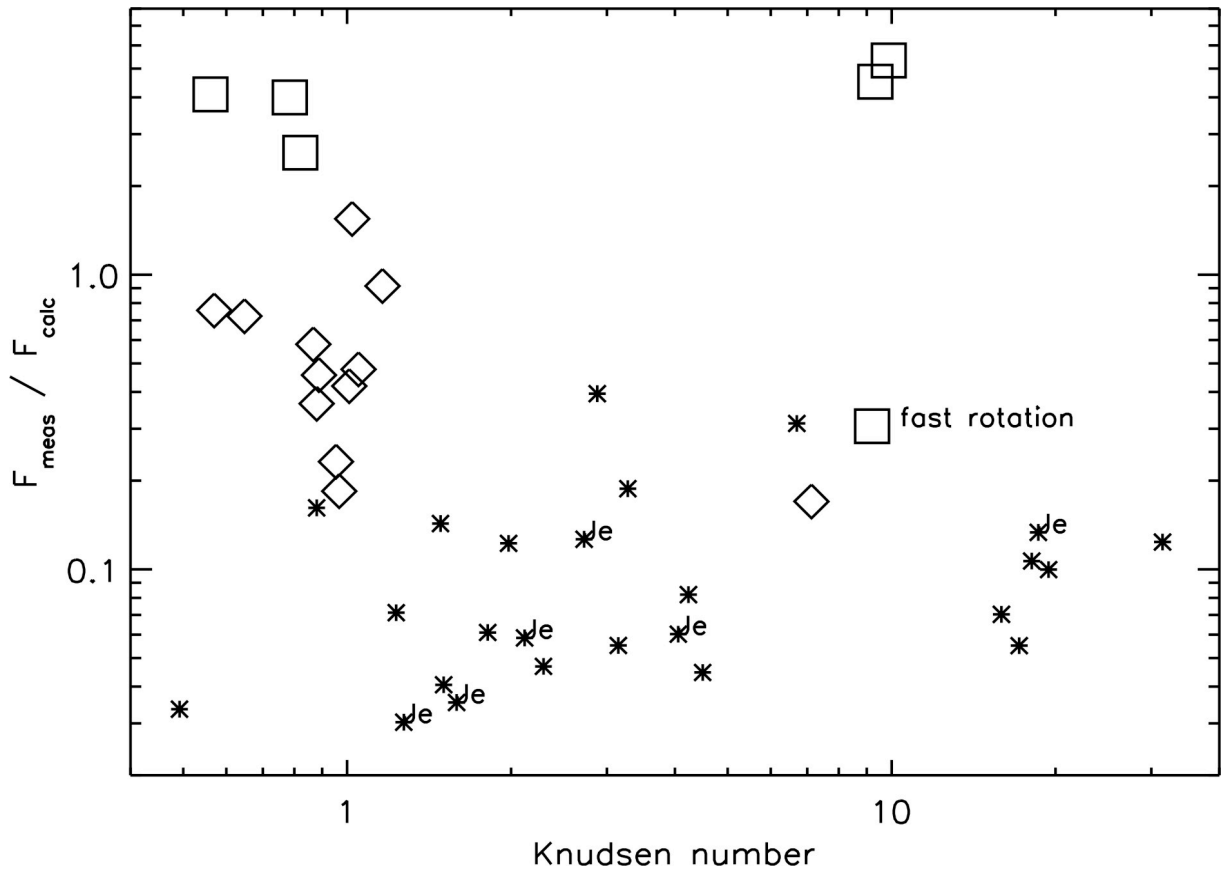

Fig. 8 The ratio between measured photophoretic force and photophoretic force calculated using eq. 1 assuming $k=1$W/mK, $J=0.5$, and $\alpha=1$. stars: chondrules, diamonds: dust mantled particles, squares: dust aggregates. To highlight the spread of on individual particle, the data from chondrule Chond-0-Je are labeled "Je".

The measured value for chondrules is below the calculated value. This can easily be explained. Non perfect absorption, especially not at the front side, a somewhat smaller accommodation coefficient and only slightly higher thermal conductivity can account for this. Dust aggregates at low pressure easily have thermal conductivities as low as k=0.01 W/mK. This alone can account for the higher values measured of dust covered particles and dust aggregates.

Fig. 8 also shows one chondrule (Chond-0-Je) which has flown several times. The variations are comparable to the total variations observed for different chondrules. Considering the systematic error of a factor of 2 by variation within the light flux and not well defined initial rotation or orientation states, it is not clear how much of the variation observed is due to variations in the individual particle properties (surface material, homogeneity, and orientation) or due to the limiting accuracy of the measurements. However, with the given accuracy the experimental data on the photophoretic strength away from the light source are in agreement to simple estimates based on spherical particles assuming typical values, i.e. typical thermal conductivities.

5. Particle rotation

In a protoplanetary disk rotation of mm-size particles might occur and has many aspects of excitation, alignment and damping. Krauss et al. (2007) argue that prolonged rotation around axes perpendicular to the direction of incident radiation is likely insignificant for mm-size particles. In that case, rotation does not change front and backside of a particle and the radial photophoretic force (away from a star) is not influenced by radiation as the temperature gradients along the particle remain constant with time. However, rotation does occur in a few cases in the experiment reported here as the damping time is much longer. Our experiments therefore allow us to make some statements on the interplay between particle rotation and photophoretic forces. As we resolve the particles spatially (see fig. 3), a rotation is immediately visible in the videos. A detailed analysis to correlate rotation to variations in photophoresis has not been attempted here as this requires a dedicated study with increased absolute accuracy. However, particles show some initial rotation, typically much slower than 2 Hz, most particles are below 1 Hz. As argued above, temperature gradients are essentially established on timescales much shorter than 1s. Therefore, the slow rotation is not significantly reducing the linear photophoretic force. There are a few exceptions. In one case a particle is spinning rapidly with about 6.3 Hz after a collision, continues to move within the light beam but shows a relatively weak photophoretic force. This particle is marked in fig. 8. For high spin rates the temperature gradient cannot fully be established along the particle before front and back change sides. Probably the weak force on the 6.3Hz particle is due to the fast spinning. Detailed analysis is beyond the scope of this paper and subject for future analysis.

Rotation typically does not just flip front and back side, but the warm front side is moving into the shadow while the cool back side is turning into the light and heats up. In this way a temperature gradient occurs in a direction perpendicular to the illumination and a photophoretic component in a direction perpendicular to the direction of illumination occurs. This is in analogy to the well known Yarkovsky effect which changes the motion of small asteroids due to the radiation pressure of thermal radiation from an asteroid on long timescales (Rubincam 1995). We also observed a sideward motion for the fast rotator here. It has to be noted though that rotation is damped on gas grain coupling timescales. So eventually, a disturbed and fast rotating particle supposedly would spin down significantly and the photophoretic force would increase along the direction of light.

Motion perpendicular to the direction of illumination also occurs for non-rotating bodies due to inhomogeneous surface heating of real particles. We did not quantify this component yet but clearly saw particles change their direction of motion correlated to a slight change in orientation. Last not least photophoresis not only reacts to rotation but also induces it. Again, due to the non-perfect particle properties (shape, internal heterogeneity) a torque can be generated which changes the particle orientation or even induces rotation. We also observed such particles which change their rotation due to photophoresis. Typically this will spin up a particle until it is balanced by gas drag. It is important to note that the spin up will usually occur around the direction of illumination, where a constant unidirectional torque can persist. A torque around an axis perpendicular to the direction of illumination will change over time as the particle reacts by rotation and as the rotation changes the geometry with respect to the illumination, e.g. a slightly elongated particle where the center of mass is not at the center of symmetry will always be subject to an aligning torque component. A more detailed treatment on particle rotation can be found in Krauss et al. (2007). We are aware that these arguments are not universal. One can construct a body which at least for a certain time span would spin up around a direction perpendicular to the direction of light but this seems to be a fairly unlikely case. Such studies are subject to further research but an example of a particle which spins up – around the direction of light – is given in fig 9.

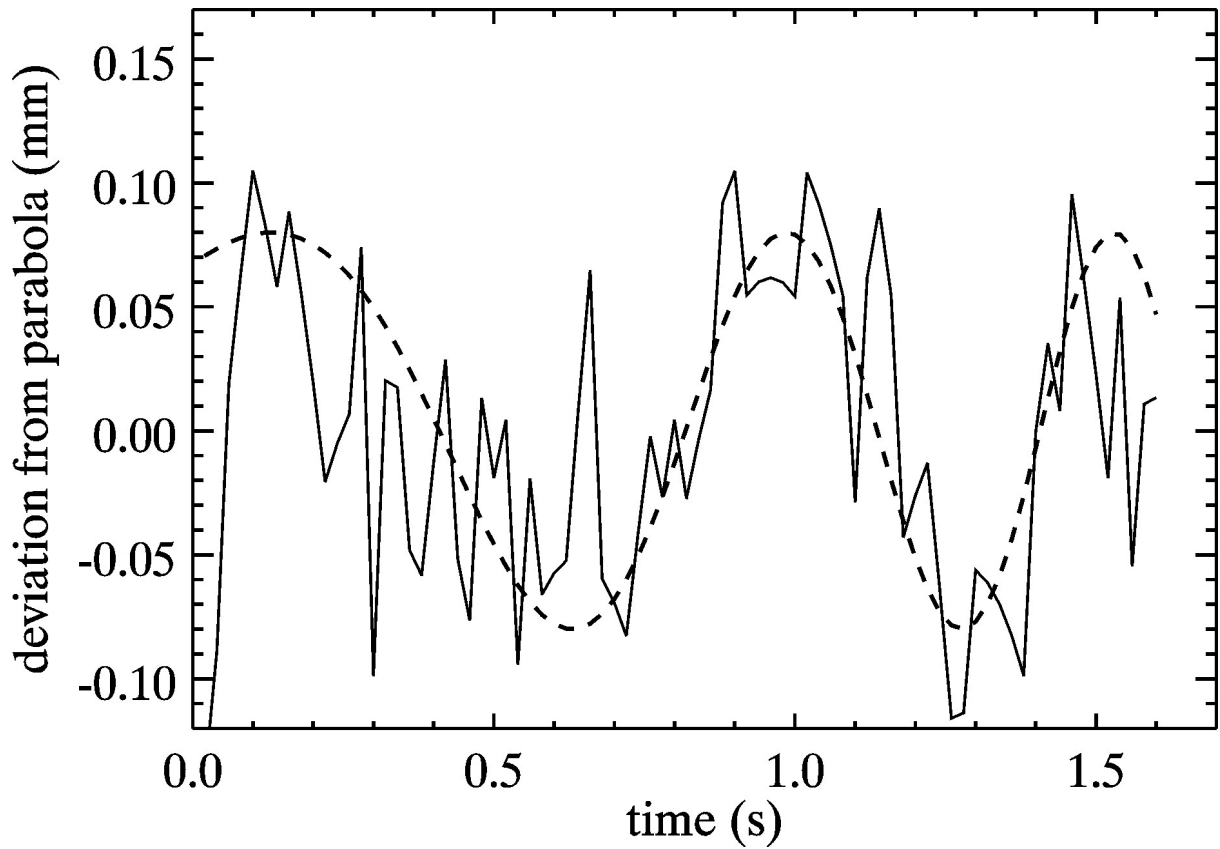

Fig. 9: Spin up of a particle around the direction of light. Shown are the deviations of a chondrule's side motion from the parabolic track and a line through the data. This is no fit but only to guide the eye.

In total, rotation up to a few Hz is not decreasing the strength of photophoresis acting on chondrules in protoplanetary disks significantly. Larger rotation frequencies change this but are not likely to occur for a long time in protoplanetary disks for a variety of reasons as pointed out in great detail in Krauss et al. (2007). More dedicated experimental studies are certainly needed here in the future.

6. Conclusions, Applications to the Solar Nebula and Protoplanetary Disks

Particle transport over several AU outward occurred in early times in the Solar Nebula. This is implied by high temperature minerals found in comet Wild 2 by the Stardust sample return mission (Zolensky et al., 2006) and is very plausible to explain the existence and properties of CAIs in meteorites (Rout et al., 2009). In a number of recent papers such transport of particles in the Solar Nebula by photophoresis was modeled on the basis of simple estimates for the photophoretic strength (Krauss and Wurm, 2005; Wurm and Krauss, 2006; Mousis et al., 2007; Krauss et al., 2007; Herrmann and Krivov, 2007; Takeuchi and Krauss, 2008; Wurm and Haack, 2009).

    No experiments existed so far for the relevant particles, i.e. chondrules, CAIs, or large dust aggregates. This is the first time that photophoresis has been measured for these particle types. We could show that the photophoretic force for dust aggregates is much larger than for

chondrules. The principle result by Wurm and Krauss (2006) that dust aggregates can be transported more efficiently than bare chondrules or dust mantled chondrules therefore holds. The photophoretic force measured for bare chondrules is on average only 10% of the photophoretic strength used by Wurm and Krauss (2006). If bare chondrules exist for a long enough time this will change possible equilibrium positions for bare chondrules in a given disk model. As gas pressure (density) enters linearly in the photophoretic force at low pressure particle concentrations can still occur but further inward. However, the disk parameters of current models of protoplanetary disks are poorly constrained. The gas density in different models can vary by orders of magnitudes as indicated in table 2.

Table 2: Parameters of the experimental particle environment compared to typical protoplanetary and transitional disks. Minimum Mass Solar Nebula refers to Hayashi et al. (1985). Parameters for TW Hya are taken from Calvet et al. (2002) and Batalha et al. (2002). The flux at 1 AU for TW Hya is based on 0.45 solar luminosities.

| Parameter | Experiment | Protoplanetary Disk Minimum Mass Solar Nebula (1 AU) | Transitional Disk TW Hya (< 4 AU) |
|---|---|---|---|
| Gas density (kg/m$^3$) | $5 \cdot 10^{-6} .. 2 \cdot 10^{-4}$ | $1.4 \cdot 10^{-6}$ | $>4 \cdot 10^{-10}$ |
| Gas temperature (K) | 300 | 300 | >80 |
| Flux (W/m$^2$) | $1.9 \cdot 10^4$ | $1.4 \cdot 10^3$ | $0.63 \cdot 10^3$ (at 1 AU) |

Table 2 shows that the experiments were carried out for temperatures, light flux, and pressures expected in protoplanetary disks but the pressure will usually be much lower in later stages of disk life when gas rich disks evolve to gas poor debris disks.

This will influence the way different kind of particles are influenced by photophoresis. Depending on the disk profile chondrules can still be concentrated at the same distance, i.e. within the asteroid belt region. In any case, even a strongly reduced photophoretic force leads to an outward drift in the inner 1 AU at pressure relevant for a minimum mass solar nebula and allows to retain high thermal conductivity particles as calculated by Wurm and Krauss (2006) in a very wide range of disk models. Drift velocities for bare chondrules would be reduced by a factor 10 compared to the calculated values and drift timescales might be correspondingly longer. The drift of 1 mm particles would then be about 2 cm/s at 1 AU (cmp. Wurm and Krauss, 2006). This would still be sufficient to transport particles by several AU over the life time of the solar nebula.

However, it is questionable if chondrules in a dusty environment stay isolated for a long time. Dust particles easily stick at low collision velocities (Blum and Wurm, 2008). Chondrules might then be dust mantled. The photophoretic force for the dust mantled chondrules measured are close to the values used in previous calculations by Wurm and Krauss (2006) and the corresponding results hold.

Particle rotation does not – in general – inhibit photophoresis but influences it and is susceptible for it. It has been speculated before that variation in photophoretic strength correlate to certain particle properties, e.g. size, density, morphology, mineralogy, chemistry, and alignment. Within the resolution of the experiments we could show that dust aggregates are transported more efficiently than bare chondrules. The data do not allow a more detailed analysis e.g. of chondrule sorting yet. This is beyond the accuracy of these very first experiments but certainly within the range of future studies. The experiments confirm that photophoresis is a viable mechanism to transport particles selectively over short and large distances in the Solar Nebula.

Acknowledgement
This work has been supported by the Deutsche Forschungsgemeinschaft as part of the research group FOR 759. Access to the drop tower was funded by the European Space Agency (ESA). We thank the two reviewers for very valuable reviews.